\documentclass[sigconf]{acmart}
\AtBeginDocument{%
  \providecommand\BibTeX{{%
    \normalfont B\kern-0.5em{\scshape i\kern-0.25em b}\kern-0.8em\TeX}}}

\copyrightyear{2021} 
\acmYear{2021} 
\setcopyright{acmcopyright}\acmConference[ICS '21]{2021 International Conference on Supercomputing}{June 14--17, 2021}{Virtual Event, USA}
\acmBooktitle{2021 International Conference on Supercomputing (ICS '21), June 14--17, 2021, Virtual Event, USA}
\acmPrice{15.00}
\acmDOI{10.1145/3447818.3461703}
\acmISBN{978-1-4503-8335-6/21/06}

\usepackage{color}
\usepackage{algorithm}
\usepackage{algpseudocode}

\let\oldReturn\Return
\renewcommand{\Return}{\State\oldReturn}
\usepackage{xspace}
\usepackage{bm}
\usepackage{graphicx}
\usepackage{pgfplots}
\pgfplotsset{compat=1.16}
\usepackage[font=normalsize]{caption}
\usepackage{subcaption}
\usepackage{tikz}
\usetikzlibrary{patterns}
\usetikzlibrary{shapes}
\usepackage[nolist]{acronym}
\usepackage{balance}

\newcommand{\ALTO}{\mbox{{\sf ALTO}}\xspace}
\newcommand{\SCAL}[1]{$#1$}
\newcommand{\MATRIX}[1]{$\textbf{#1}$}
\newcommand{\MATELEM}[3]{${#1}_{#2,#3}$}
\newcommand{\VECTOR}[1]{$\textbf{#1}$}
\newcommand{\VECELEM}[2]{${#1}_{#2}$}
\newcommand{\TENSOR}[1]{$\mathcal{#1}$}
\newcommand{\TENSORI}[1]{\mathcal{#1}}
\newcommand{\TENELEM}[3]{$#1_{#2_{1},#2_{2},\dots,#2_{#3}}$}
\newcommand{\MATTEN}[2]{$\textbf{#1}_{(#2)}$}

\newcommand{\REALTWO}[2]{$\mathbb{R}^{#1\times #2}$}
\newcommand{\REALTHREE}[3]{$\mathbb{R}^{#1\times #2 \times #3}$}
\newcommand{\REALX}[2]{$\mathbb{R}^{#1_{1}\times #1_{2}\times \cdots \times #1_{#2}}$}

\newcommand{\IGNORE}[1]{}
\newcommand{\REALIJK}{$\mathbb{R}^{I\times J\times K}$}

\newcommand{\second}{\mbox{s}}

\newcommand{\GBS}{\mbox{GB/\second}}

\newcommand{\GHZ}{\mbox{GHz}}

\newcommand{\rl}{roof{}line}
\newcommand{\Rlm}{Roof{}line model}

\newcommand*\bcircled[1]{\tikz[baseline=(char.base)]{\node[shape=circle,fill,inner sep=0.5pt] (char) {\small \textcolor{white}{#1}};}}

\begin{acronym}[MTTKRP]
\acro{CLX}{Cascade Lake-X}
\acro{FROSTT}{the Formidable Repository of Open Sparse Tensors and Tools}
\acro{HaTen2}{Hadoop Tensor method for 2 decompositions}
\acro{ICPC}{Intel C++ Compiler}
\acro{MTTKRP}{matricized tensor times Khatri-Rao product}
\end{acronym}

\definecolor{greyback}{RGB}{248,248,248}
\definecolor{deepblue}{RGB}{43,131,186}
\definecolor{greenalt}{RGB}{35,139,69}
\definecolor{darkred}{RGB}{215,25,28}
\definecolor{darkorange}{RGB}{253,174,97}
\definecolor{light-gray}{gray}{0.85}
\definecolor{lighter-gray}{gray}{0.95}
\definecolor{lighter-gray2}{gray}{0.98}

\begin{document}

\title{ALTO: Adaptive Linearized Storage of Sparse Tensors}

\author{Ahmed E. Helal}
\affiliation{\institution{Intel Labs}\country{}}
\email{ahmed.helal@intel.com}

\author{Jan Laukemann}
\affiliation{\institution{Intel Labs}\country{}}
\email{jan.laukemann@intel.com}

\author{Fabio Checconi}
\affiliation{\institution{Intel Labs}\country{}}
\email{fabio.checconi@intel.com}

\author{Jesmin Jahan Tithi}
\affiliation{\institution{Intel Labs}\country{}}
\email{jesmin.jahan.tithi@intel.com}

\author{\hsize=0.3\textwidth Teresa Ranadive}
\affiliation{\institution{Laboratory for Physical Sciences}\country{}}
\email{tranadive@lps.umd.edu}

\author{Fabrizio Petrini}
\affiliation{\institution{Intel Labs}\country{}}
\email{fabrizio.petrini@intel.com}

\author{Jeewhan Choi}
\affiliation{\institution{University of Oregon}\country{}}
\email{jeec@uoregon.edu}

\renewcommand{\shortauthors}{Ahmed E. Helal, Jan Laukemann, Fabio Checconi, et al. }

\begin{abstract}
The analysis of high-dimensional sparse data is becoming increasingly popular in many important domains. However, real-world sparse tensors are challenging to process due to their irregular shapes and data distributions. We propose the Adaptive Linearized Tensor Order (ALTO) format, a novel mode-agnostic (general) representation that keeps neighboring nonzero elements in the multi-dimensional space close to each other in memory. To generate the indexing metadata, \ALTO uses an adaptive bit encoding scheme that trades off index computations for lower memory usage and more effective use of memory bandwidth. Moreover, by decoupling its sparse representation from the irregular spatial distribution of nonzero elements, \ALTO eliminates the workload imbalance and greatly reduces the synchronization overhead of tensor computations. As a result, the parallel performance of \ALTO-based tensor operations becomes a function of their inherent data reuse. On a gamut of tensor datasets, \ALTO outperforms an oracle that selects the best state-of-the-art format for each dataset, when used in key tensor decomposition operations. Specifically, \ALTO achieves a geometric mean speedup of $8\times$ over the best mode-agnostic (coordinate and hierarchical coordinate) formats, while delivering a geometric mean compression ratio of $4.3\times$ relative to the best mode-specific (compressed sparse fiber) formats.
\end{abstract}



\maketitle

\section{Introduction}
Many critical application domains such as healthcare~\cite{ho2014marble, wang2015rubik}, cybersecurity~\cite{kobayashi2018extracting, fanaee2016tensor}, data mining~\cite{kolda2008scalable, papalexakis2016tensors}, and machine learning~\cite{anandkumar2014tensor, sidiropoulos2017tensor} produce and manipulate massive amounts of high-dimensional data. Such datasets can naturally be represented as sparse tensors, which store the values of the nonzero tensor elements along with indexing metadata that denote the position of each nonzero in the tensor.
Therefore, a fundamental problem in sparse tensor computations involves determining how to store, group, and organize the nonzero elements to 1)~reduce memory storage, 2)~improve data locality, 3)~increase parallelism, and 4)~decrease workload imbalance and synchronization overhead. 
Since tensor algorithms perform computations along different mode (i.e., dimension) orientations, practical sparse tensor formats must be mode-agnostic to deliver acceptable performance and scalability across all modes. Because real-world sparse tensors are highly irregular in terms of their shape, dimensions, and distribution of nonzero elements, achieving these (oftentimes conflicting) goals is challenging.

To tackle this problem, researchers have proposed many sparse tensor formats~\cite{Bader2007, baskaran2012efficient, Smith2015, Smith2015a, liu2017unified, phipps2019software, li2018hicoo, li2019efficient, nisa2019load, nisa2019efficient}, which can be classified based on their encoding of the multi-dimensional coordinates into list-, block-, and tree-based formats~\cite{chou2018format}. 
List-based tensor representations, such as the simple coordinate~(COO) format, explicitly store the nonzero elements along with their coordinates (i.e., the indices of all dimensions). Therefore, they are agnostic to the different mode orientations of tensor algorithms and, as a result, they remain the de facto sparse tensor storage~\cite{chou2018format} in many libraries (e.g., Tensor Toolbox~\cite{Bader2007}, Tensorflow~\cite{abadi2016tensorflow}, and Tensorlab~\cite{vervliet2016tensorlab}). However, the list-based COO format does not impose any order on the multi-dimensional data and it suffers from a significant synchronization overhead to resolve the update/write conflicts across threads~\cite{liu2017unified}.

Prior block-based sparse tensor representations employ multi-dimensional tiling schemes to further compress the COO format~\cite{li2018hicoo, li2019efficient}. However, the efficacy of this hierarchical COO~(HiCOO) storage completely depends on the characteristics of target tensors (such as their shape, density, and data distribution) as well as the format parameters (e.g., block size).
In addition, the resulting parallel schedule of HiCOO blocks can suffer from limited parallelism and scalability due to conflicting updates across blocks. 

\begin{figure*}[htb]
\centering
\includegraphics[trim=55 345 55 343, clip, width=0.98\linewidth]{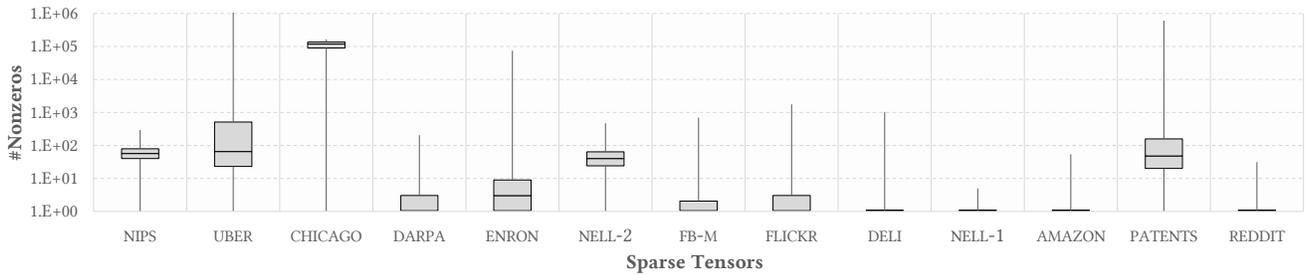}
	\vspace*{-10pt}
	\caption{A box plot of the data (nonzero elements) distribution across the multi-dimensional blocks (subspaces) of the hierarchical coordinate storage~\cite{li2018hicoo}.
	The multi-dimensional subspace size is 128$^{N}$, where $N$ is the number of dimensions (modes), as per prior work~\cite{sun2020sptfs}. The sparse tensors are sorted in an increasing order of their size (number of nonzero elements).}
	\vspace*{-6pt}
\label{fig:motivation}
\end{figure*}

Several proposals use tree-based data structures to extend traditional compressed matrix formats, such as the compressed sparse row~(CSR) format, to higher-order tensors. The most popular example of these storage representations is the compressed sparse fiber~(CSF) format~\cite{Smith2015}, which uses multiple arrays of index pointers to compress the multi-dimensional indices of nonzero elements. While CSF-based formats can reduce memory traffic, they are mode-specific, i.e., they are oriented to favor a specific order of tensor modes. 
As the tensor order increases, these mode-specific formats require excessive memory to store multiple tensor copies~\cite{smith2017accelerating} and/or different code implementations for computing along distinct modes~\cite{Smith2015a}.
Moreover, such a compressed, coarse-grained format can suffer from significant workload imbalance and limited scalability, especially in irregularly shaped tensors with short modes~\cite{li2018hicoo}. 

In summary, the above block- and tree-based approaches for sparse tensor storage rely on \textit{clustering the nonzero elements based on their location in the multi-dimensional space} and then partitioning this space into \textit{non-overlapping regions} to generate a compressed indexing metadata. Hence, they are constrained by the spatial distribution of nonzero elements. 
To this end, Figure~\ref{fig:motivation} shows the distribution of nonzero elements in the multi-dimensional space for 
a set of 3D and 4D sparse tensors.\footnote{The implementation of HiCOO~\cite{li2018hicoo}, a block-based COO format, only supports 3D and 4D sparse tensors and it does not auto-tune the blocking sizes. We include ten datasets from~\cite{li2018hicoo}, and add three large-scale datasets with billions of elements. } Specifically, it demonstrates that
the number of nonzero elements per block fluctuates widely (note the use of logarithmic scale). As the sparsity of tensors increases (e.g., \textsc{deli}, \textsc{nell-1}, \textsc{amazon}, and \textsc{reddit} tensors), the location-based clustering fails to compress tensors and introduces a substantial memory overhead. As a result, the parallel performance and scalability of such location-based formats can be severely impacted by the irregular data distributions and unstructured sparsity patterns that typically emerge in higher-order sparse tensors.   

\vspace*{-5pt}
\subsection{Adaptive Linearized Tensor Order}
We propose the \textit{Adaptive Linearized Tensor Order (\ALTO)} format, a mode-agnostic storage system for sparse tensors that addresses the irregularity in sparse tensor computations and the performance bottlenecks on modern parallel architectures. \ALTO organizes and stores the nonzero elements of a given tensor in a one-dimensional data structure along with compact indexing metadata, such that neighboring nonzero elements in the tensor are close to each other in memory.
Most importantly, it generates the indexing metadata using an adaptive bit encoding scheme based on the shape and dimensions of target tensors. Such an adaptive format trades off index computations (i.e., de-linearization) for lower memory usage and more efficient use of the effective memory bandwidth. 

Unlike prior compressed~\cite{Smith2015, li2018hicoo} and linearized~\cite{harrison2018high} sparse tensor formats, \ALTO not only improves data locality across all mode orientations but also \textit{eliminates workload imbalance} and greatly reduces synchronization cost, which have traditionally limited the performance and scalability of irregular sparse tensor computations. Moreover, it generates perfectly balanced partitions for effective parallel execution. 
Although each nonzero tensor element is strictly mapped to one partition, the subspace coordinates of \ALTO partitions may overlap. 
To resolve update conflicts across these potentially overlapping partitions, \ALTO locates their positions in the multi-dimensional space and automatically selects the appropriate synchronization mechanism based on the data reuse of target tensors. As a result, \ALTO delivers superior performance over state-of-the-art sparse tensor formats. 
In what follows, we summarize the contributions of this work:
\begin{itemize}
\item We introduce \ALTO, a novel sparse tensor format for high-performance and scalable execution of tensor operations. Unlike prior location-based clustering approaches for compressed sparse tensor storage, \ALTO uses a single (mode-agnostic) tensor representation that improves data locality, eliminates workload imbalance, and greatly reduces memory usage and synchronization overhead, \textit{regardless of the data distribution in the multi-dimensional space} ($\S$\ref{sec:approach}).
\item We propose an adaptive synchronization mechanism to efficiently resolve conflicting updates (writes) across threads based on the average data reuse of target sparse tensors ($\S$\ref{sec:approach}).
\item We present effective parallel algorithms to perform Matricized Tensor Times Khatri-Rao Product (MTTKRP) operations, the main tensor decomposition performance bottleneck~\cite{Smith2015, baskaran2012efficient, kang2012gigatensor, Choi2014}, on sparse tensors stored via \ALTO ($\S$\ref{sec:approach}). 
\item We demonstrate that \ALTO outperforms the state-of-the-art sparse tensor formats via experimental evaluation over a variety of real-world datasets. \ALTO achieves a geometric mean speedup of $8\times$ over the best mode-agnostic format and a geometric mean compression ratio of $4.3\times$ over the best mode-specific format ($\S$\ref{sec:perf}).
\end{itemize} 

\section{Background}\label{sec:background}
We begin with a brief overview of tensor decomposition methods and related notations.
The work by Kolda and Bader~\cite{Kolda2009,Bader2007} provides a more in-depth discussion of tensor algorithms and applications.

\subsection{Notations}
Tensors are the higher-order generalization of matrices.
An $N$ dimensional tensor is said to have $N$ \emph{modes} and is called a mode-$N$ tensor.
The following notations are used in this paper:
\begin{enumerate}
	\item
		\emph{Scalars} are denoted by lower case letters (e.g., \SCAL{a}).
	\item
		\emph{Vectors} are mode-$1$ tensors denoted by bold
		lower case letters (e.g., \VECTOR{a}). The $i^{th}$ element of a
		vector \VECTOR{a} is denoted by \VECELEM{a}{i}.
	\item
		\emph{Matrices} are mode-$2$ tensors denoted by bold
		capital letters (e.g., \MATRIX{A}). If \MATRIX{A} is a $I\times J$ matrix,
		it can also be denoted as \MATRIX{A}~$\in$~\REALTWO{I}{J},
		and its element at index $(i,j)$ is denoted as \MATELEM{a}{i}{j}.
	\item
		\emph{Higher-order tensors} are denoted by Euler script letters (e.g., \TENSOR{X}).
		A mode-$N$ tensor whose dimensions are $I_{1}\times I_{2}\times \cdots \times I_{N}$
		can be denoted as \TENSOR{X} $\in$ \REALX{I}{N},
		and its element at index $(i_{1},i_{2},\dots,i_{N})$ is denoted as \TENELEM{x}{i}{N}.
	\item
		\emph{Fibers} are the higher-order analogue of matrix rows and columns.
		A mode-$n$ fiber is defined by fixing every mode except the $n^{th}$ mode. 
		For example, a matrix column is defined by fixing the second
		mode, and is therefore a mode-$1$ fiber.
		A fiber is denoted by using a colon for the non-fixed
		mode (e.g., the $j^{th}$ column of a matrix A is denoted by \VECTOR{a}$_{:,j}$).
		\item \emph{Slices} are the lower-order components of a tensor which result from fixing all but two modes. For example, the slices of a mode-3 tensor \TENSOR{X} are matrices (such as \TENSOR{X}$_{i,:,:}$ and \TENSOR{X}$_{:,j,:}$). 
\end{enumerate}

\subsection{Canonical Polyadic Decomposition}\label{sec:cpd-als}
The Canonical Polyadic Decomposition (CPD) is a widely used type of tensor factorization, in which a mode-$N$ tensor \TENSOR{X} is approximated by the sum of $R$ outer products of $N$ vectors.  Each outer product is called a \emph{rank-$1$ component}, while the sum of the $R$ components is said to be a \emph{rank-$R$ decomposition} of \TENSOR{X}. 
The vectors forming the rank-$1$ outer products each correspond to a particular tensor mode.  We may arrange the $R$ vectors corresponding to each of the $N$ modes into $N$ different \emph{factor matrices} so that the decomposition of \TENSOR{X} is the outer product of these matrices.  
For example, if \TENSOR{X}~$\in$~\REALTHREE{I}{J}{K}, we may write a decomposition of \TENSOR{X} in terms of factor matrices \MATRIX{A}~$\in$~\REALTWO{I}{R}, \MATRIX{B}~$\in$~\REALTWO{J}{R}, and \MATRIX{C}~$\in$~\REALTWO{K}{R}, where
the columns of \MATRIX{A} (resp. \MATRIX{B} and \MATRIX{C}) are the vectors used in forming the $R$ outer products along mode-$1$ (resp. $2$ and $3$).

The CPD-Alternating Least Squares~(CPD-ALS) method is a popular tensor decomposition algorithm.
During each ALS iteration, one alternates between updating each of the individual factor matrices, i.e., updating a factor matrix to yield the best approximation of \TENSOR{X} when all other factor matrices are fixed.
The most expensive part of CPD-ALS, along with many other tensor algorithms, is the \ac{MTTKRP} operation~\cite{Smith2015}. 

The MTTKRP operation involves two basic subroutines: 
\begin{enumerate}
	\item 
	\emph{Tensor matricization} -- a process where a tensor is unfolded or flattened into a matrix. Moreover, the mode-$n$ matricization of a tensor \TENSOR{X}, denoted \MATTEN{X}{n}, is obtained by laying out the mode-$n$ fibers of \TENSOR{X} as the columns of \MATTEN{X}{n}.
	Hence, when \MATTEN{X}{n} is multiplied by the Khatri-Rao product (see below), the tensor indices associated with the $q$-th column of \MATTEN{X}{n} match those given by the rows of the factor matrices used to form the $q$-th row of the Khatri-Rao product.
	\item
	\emph{Khatri-Rao product}~\cite{KR_product} -- the ``matching column-wise'' Kronecker product between two matrices.
	That is, given matrices 
	\MATRIX{B} $\in$ \REALTWO{J}{R}
	and \MATRIX{C} $\in$ \REALTWO{K}{R},
	their Khatri-Rao product \MATRIX{K},
	denoted \MATRIX{K} $=$ \MATRIX{B} $\odot$ \MATRIX{C},
	where \MATRIX{K} is a $(J\cdot K) \times R$ matrix, is defined as:	
	$B\odot C = \left[\textbf{b}_{1}\otimes \textbf{c}_{1}\ \textbf{b}_{2}\otimes \textbf{c}_{2} \dots \textbf{b}_{R}\otimes \textbf{c}_{R}\right] \nonumber$.
\end{enumerate}

For a mode-$3$ tensor \TENSOR{X}, the mode-1 MTTKRP operation can be expressed as
$\textbf{X}_{(1)} \left(\textbf{B}\odot \textbf{C}\right)$. Typically, MTTKRP operations along all modes are performed 10--100 times in one tensor decomposition calculation. Since these MTTKRP operations are similar, we only discuss mode-$1$ MTTKRP in this paper.

\section{ALTO Format}\label{sec:approach}
Real-world sparse tensors, which emerge in high-dimensional data analytics, are challenging to efficiently encode and represent as they suffer from highly irregular shapes and data distributions as well as unstructured sparsity patterns. For example, one mode of a tensor may represent a massive user database while another mode represents their demographic information, their interactions, or their (potentially incomplete) consumer preferences/ratings for a set of products~\cite{frostt}.

Thus, the proposed \ALTO format uses an adaptive (data-aware) recursive partitioning of the high-dimensional space that represents a given sparse tensor to generate a mode-agnostic linearized index, which maps a point (nonzero element) in this Cartesian space to a point on a compact line. Specifically, \ALTO splits every mode into multiple regions based on the mode length, such that each distinct mode has a variable number of regions to adapt to the unequal cardinalities of different modes and to minimize the storage requirements.
This adaptive linearization and recursive partitioning of the multi-dimensional space ensures that neighboring points in space are close to each other on the resulting compact line, thereby maintaining the inherent data locality of tensor algorithms. 
Moreover, the \ALTO format is not only locality-friendly, but also parallelism-friendly as it decomposes the multi-dimensional space into perfectly balanced (in terms of workload) subspaces.
Further, it intelligently arranges the modes in the derived subspaces based on their cardinality (dimension length) to further reduce the overhead of resolving the write conflicts that typically occur in parallel sparse tensor computations.

What follows is a detailed description and discussion of the \ALTO format generation ($\S$\ref{sec:approach-gen}) and the workload partitioning and scheduling methods ($\S$\ref{sec:approach-par}) using a concrete walk-through example. In addition, we present the \ALTO-based sequential and parallel algorithms as well as adaptive conflict resolution mechanisms for efficient execution of illustrative sparse tensor operations on parallel shared-memory platforms ($\S$\ref{sec:approach-ops}). 

\subsection{ALTO Tensor Generation}\label{sec:approach-gen}
Formally, an \ALTO tensor \TENSOR{X} $= \{$\VECELEM{x}{1}, \VECELEM{x}{2}, $\dots$, \VECELEM{x}{M}$\} $ is an ordered set of nonzero elements, where each element \VECELEM{x}{i} $= \langle v_{i}, p_{i} \rangle $ is represented by a value $v_{i}$ and a position $p_{i}$. 
The position $p_{i}$ corresponds to a compact mode-agnostic encoding of the indexing metadata, which is used to quickly generate the tuple $(i_{1},i_{2},\dots,i_{N})$ that locates a nonzero element in the multi-dimensional Cartesian space.        

The generation of an \ALTO tensor is carried out in two stages: linearization and ordering. First, \ALTO constructs the indexing metadata using a compressed encoding scheme, based on the cardinalities of tensor modes, to map each nonzero element to a position on a compact line. Second, it arranges the nonzero elements in linearized storage according to their line positions, i.e., the values of their \ALTO index. 
Typically, the ordering stage dominates the format conversion/generation time. However, compared to the location-based sparse tensor formats~\cite{Smith2015, Smith2015a, li2018hicoo, li2019efficient, nisa2019load, nisa2019efficient}, \ALTO requires a minimal generation time because ordering the linearized tensors incurs a fraction of the cost required to sort the multi-dimensional tensor formats (due to the reduction in the comparison operations, as detailed in $\S$\ref{sec:perf}).

\begin{figure}[tb]
\centering
\includegraphics[width=0.99\linewidth]{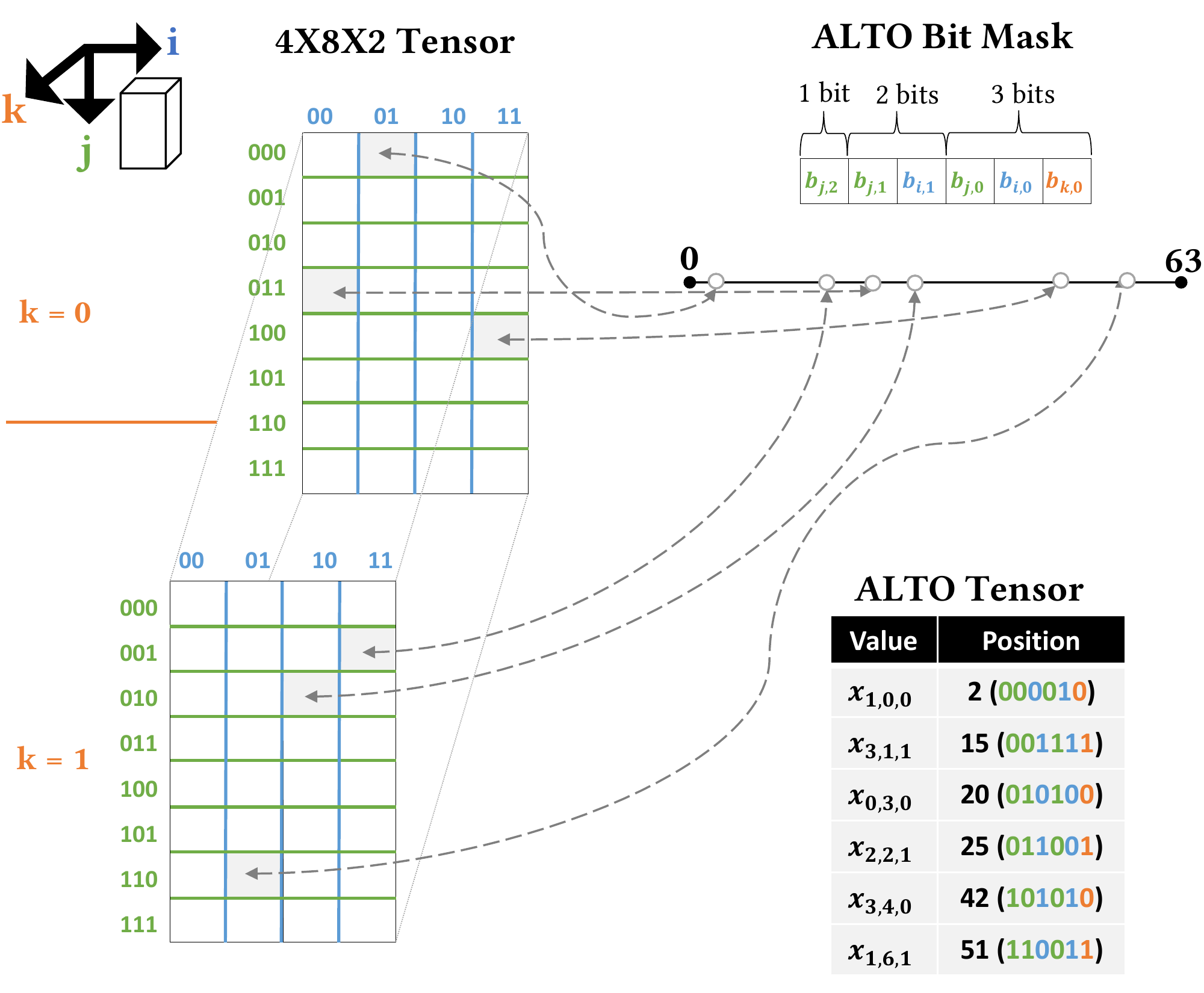}
	\vspace*{-14pt}
	\caption{An example of the \ALTO sparse encoding and representation for a three-dimensional tensor.}
	\vspace*{-13pt}
\label{fig:alto_encoding}
\end{figure}

Figure~\ref{fig:alto_encoding} provides an example of the \ALTO format for a $4\times8\times2$ sparse tensor with six nonzeros (denoted $x_{i,j,k}$). The multi-dimensional indices ($i$, $j$, and $k$) are color coded and the $r^{th}$ bit of their binary representation is denoted $b_{i/j/k, r}$. 
Specifically, \ALTO keeps the value of a nonzero element along with a linearized index, where each bit of this index is selected to partition the multi-dimensional space into two hyperplanes.
For example, the \ALTO encoding in Figure~\ref{fig:alto_encoding} uses a compact line of length $64$ (i.e., a $6$-bit linearized index) to represent the target tensor of size $4\times8\times2$. This index consists of three groups of bits with variable sizes (resolutions) to efficiently handle high-order data of arbitrary dimensionality. Within each bit group, \ALTO arranges the modes in increasing order of their length (i.e., the shortest mode first), which is equivalent to partitioning the multi-dimensional space along the longest mode first. Such an encoding aims to generate a balanced linearization of the irregular Cartesian space, where \textit{the amount of information about the spatial position of a nonzero element decreases with every consecutive bit}, starting from the most significant bit. Therefore, the line segments encode subspaces with mode intervals of equivalent lengths, e.g., the line segments $[0-31]$, $[0-15]$, and $[0-7]$ encode subspaces of $4\times4\times2$, $4\times2\times2$, and $2\times2\times2$ dimensions, respectively.   

Due to this adaptive partitioning of the multi-dimensional data, \ALTO encodes the resulting linearized index in the minimum number of bits, and it improves data locality across all modes of a given sparse tensor. Hence, a mode-$N$ tensor, whose dimensions are $I_{1}\times I_{2}\times \cdots \times I_{N}$, can be efficiently represented using a single \ALTO format with indexing metadata of size:
\vspace*{-4pt}
\begin{equation}
    S_{\text{ALTO}} = M \times (\sum_{n=1}^{N} \log_2 I_{n}) \text{ bits,}
\end{equation}
where $M$ is the number of nonzero elements.

As a result, compared to the de facto COO format, \ALTO reduces the storage requirement regardless of the tensor's characteristics. That is, the metadata compression ratio of the \ALTO format relative to COO is always $\ge 1$. On a hardware architecture with a word-level memory addressing mode, this compression ratio is given by:
\begin{equation}
    \frac{S_{\text{COO}}}{S_{\text{ALTO}}} = \frac{\sum_{n=1}^{N} \Big\lceil\frac{\log_2 I_{n}}{W_b}\Big\rceil}{\bigg\lceil\frac{\sum_{n=1}^{N} \log_2 I_{n}}{W_b}\bigg\rceil},
\end{equation}
where $W_b$ is the word size in bits. For example, on an architecture with a byte-level addressing mode (i.e., $W_b = 8$ bits), the sparse tensor in Figure~\ref{fig:alto_encoding} requires three bytes to store the mode indices of a nonzero element in the \textit{list-based} COO format and a single byte to store the linearized index of the same element in the ALTO format: the metadata compression ratio of \ALTO, compared to the list-based formats, is three.

Most importantly, the \ALTO format not only reduces the overall memory traffic of sparse tensor computations, but also decreases the number of memory transactions required to access the indexing metadata of a sparse tensor, as reading the linearized index requires fewer memory transactions compared to reading several multi-dimensional indices. In addition, this natural coalescing of the multi-dimensional indices into a single linearized index increases the memory transaction size to make more efficient use of the main memory bandwidth.  

It is important to note that \ALTO uses a non-fractal\footnote{A fractal pattern is a hierarchically self-similar pattern that looks the same at increasingly smaller scales.} encoding scheme, unlike the traditional space-filling curves (SFCs)~\cite{peano1890courbe}. In contrast, such SFCs (e.g., Z-Morton order~\cite{morton1966computer}) are based on continuous self-similar (fractal) functions that can be extremely inefficient to encode the irregularly shaped multi-dimensional spaces that emerge in higher-order sparse tensor algorithms, as they require indexing metadata of size:
\vspace*{-2pt}
\begin{equation}
    S_{\text{SFC}} = M \times N \times \max_{n=1}^{N} (\log_2 I_{n}) \text{ bits.}
\end{equation}

\begin{figure}[tb]
\centering
\includegraphics[width=1.0\linewidth]{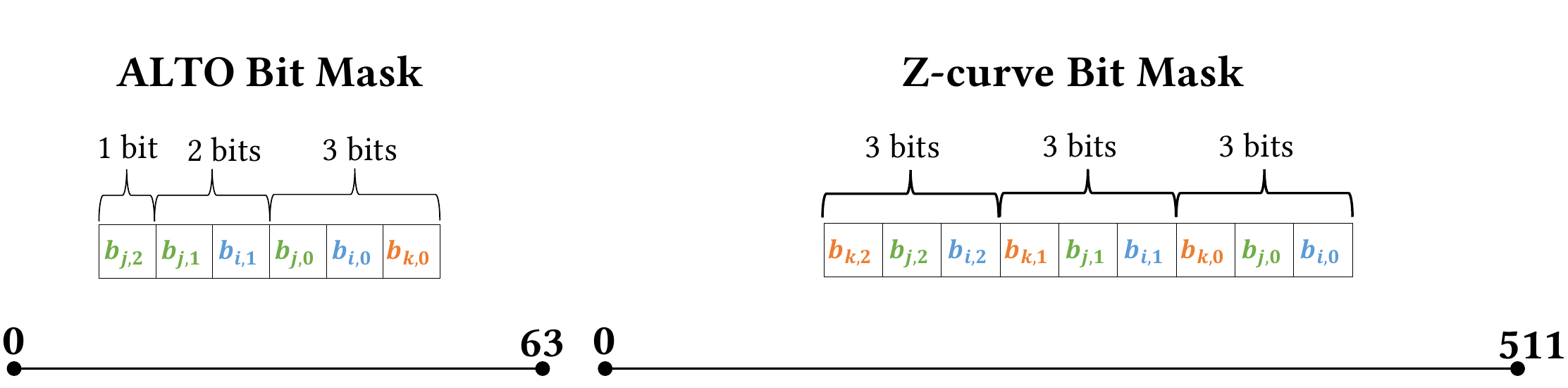}
	\vspace*{-18pt}
	\caption{For the example in Figure~\ref{fig:alto_encoding}, \ALTO generates a non-fractal, yet more compact encoding compared to traditional space-filling curves, such as the Z-Morton order.}
		\vspace*{-5pt}
\label{fig:alto_sfc}
\end{figure}

Therefore, the application of SFCs in sparse tensor computations has been \textit{limited to reordering the nonzero elements to improve their data locality} rather than compressing the indexing metadata~\cite{li2018hicoo}. Figure~\ref{fig:alto_sfc} shows the compact encoding generated by \ALTO compared to the fractal encoding scheme based on the Z-Morton curve. In this example, the non-fractal encoding scheme used by \ALTO reduces the length of the encoding line by a factor of eight, which not only decreases the overall size of the indexing metadata, but also reduces the linearization/de-linearization time required to map the multi-dimensional space to/from the compact encoding line.

\begin{figure}[tb]
\centering
\subfloat[\ALTO generates its mode-agnostic linearized index using bit-level gather operations.]{
\includegraphics[width=1.0\linewidth]{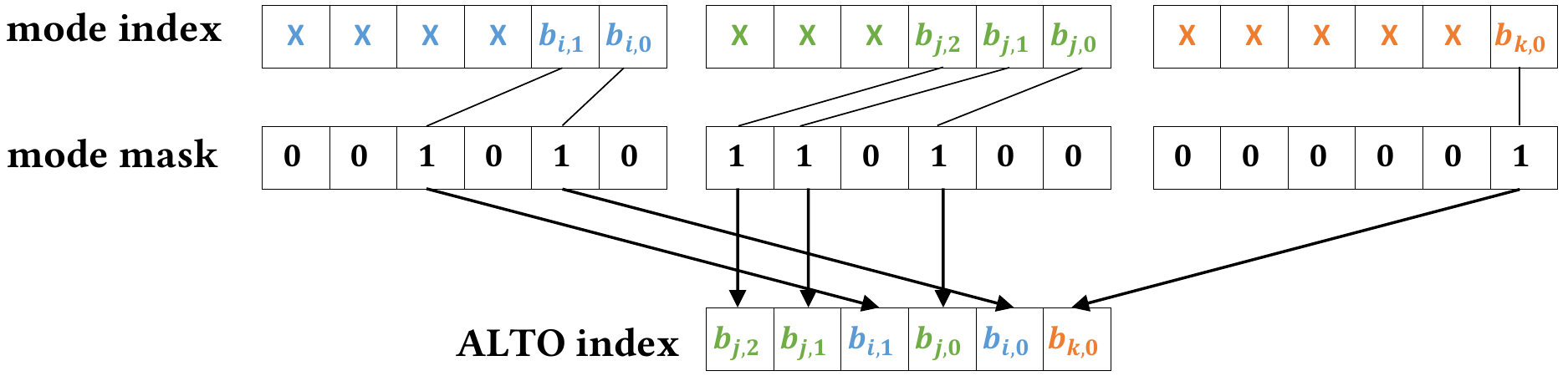}
}\vspace*{10pt}

\subfloat[To generate the multi-dimensional indices, \ALTO decodes the linearized indexing metadata using bit-level scatter operations.]{
\includegraphics[width=1.0\linewidth]{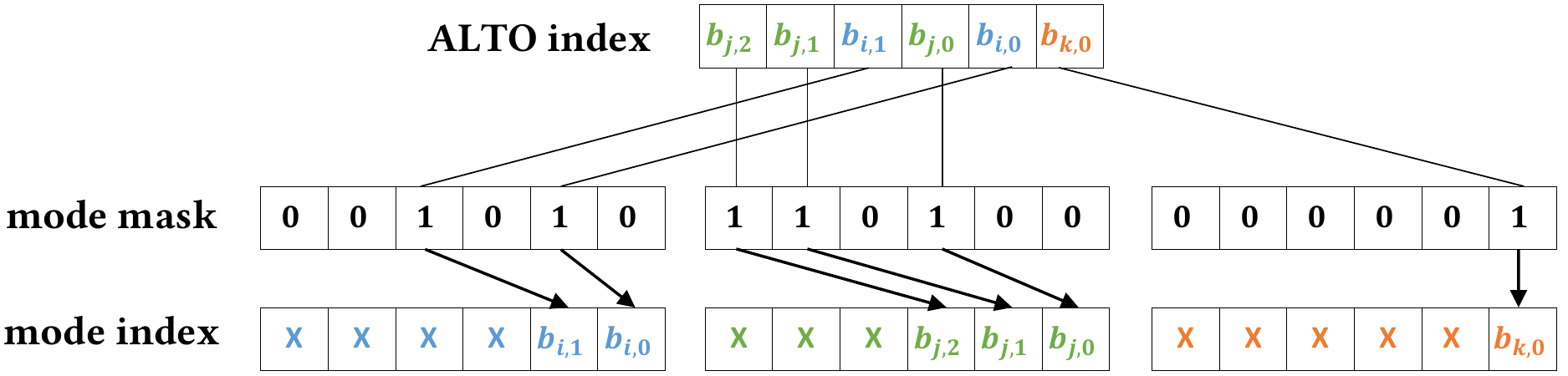}
}
	\vspace*{-7pt}
\caption{The \ALTO-based bit encoding and decoding mechanisms for the example in Figure~\ref{fig:alto_encoding}.}
	\vspace*{-3pt}
\label{fig:alto_linear}
\end{figure}

To allow fast indexing of the linearized tensors during sparse tensor operations, the \ALTO encoding is implemented using a set of simple $N$ bit masks, where $N$ is the number of modes, on top of common data processing primitives. Figure~\ref{fig:alto_linear} shows the linearization and de-linearization mechanisms, which are used during the \ALTO format generation and the sparse tensor computations, respectively. The linearization is implemented as a bit-level gather, while the de-linearization is performed as a bit-level scatter. Thus, while the compressed representation of the proposed \ALTO format comes at the cost of a de-linearization (decompression) overhead, such a computational overhead is minimal and can be effectively overlapped with the memory accesses of the memory-intensive sparse tensor operations, as shown in $\S$\ref{sec:perf}. 

\subsection{Workload Partitioning and Scheduling}\label{sec:approach-par}
The prior compressed sparse tensor formats, such as block- and CSF-based approaches, seek to reduce the size of the indexing metadata by clustering the nonzero elements into coarse-grained structures (e.g., tensor blocks, slices, and/or fibers) that divide the multi-dimensional space of a given tensor into non-overlapping regions. However, due to the irregular shapes and distributions of higher-order data, such coarse-grained approaches can suffer from severe workload imbalance, in terms of nonzero elements, which in turn leads to limited parallel performance and scalability.  

Thus, the proposed \ALTO representation works at the finest granularity level (i.e., a single nonzero element), which exposes the maximum fine-grained parallelism and allows scalable parallel execution. While a non-overlapping space partitioning of a tensor can be obtained from the \ALTO encoding scheme, using a subset of the index bits, the workload balance of such a partitioning still depends on the sparsity patterns of the tensor. 

\begin{figure}[tb]
\centering
\includegraphics[width=1.0\linewidth]{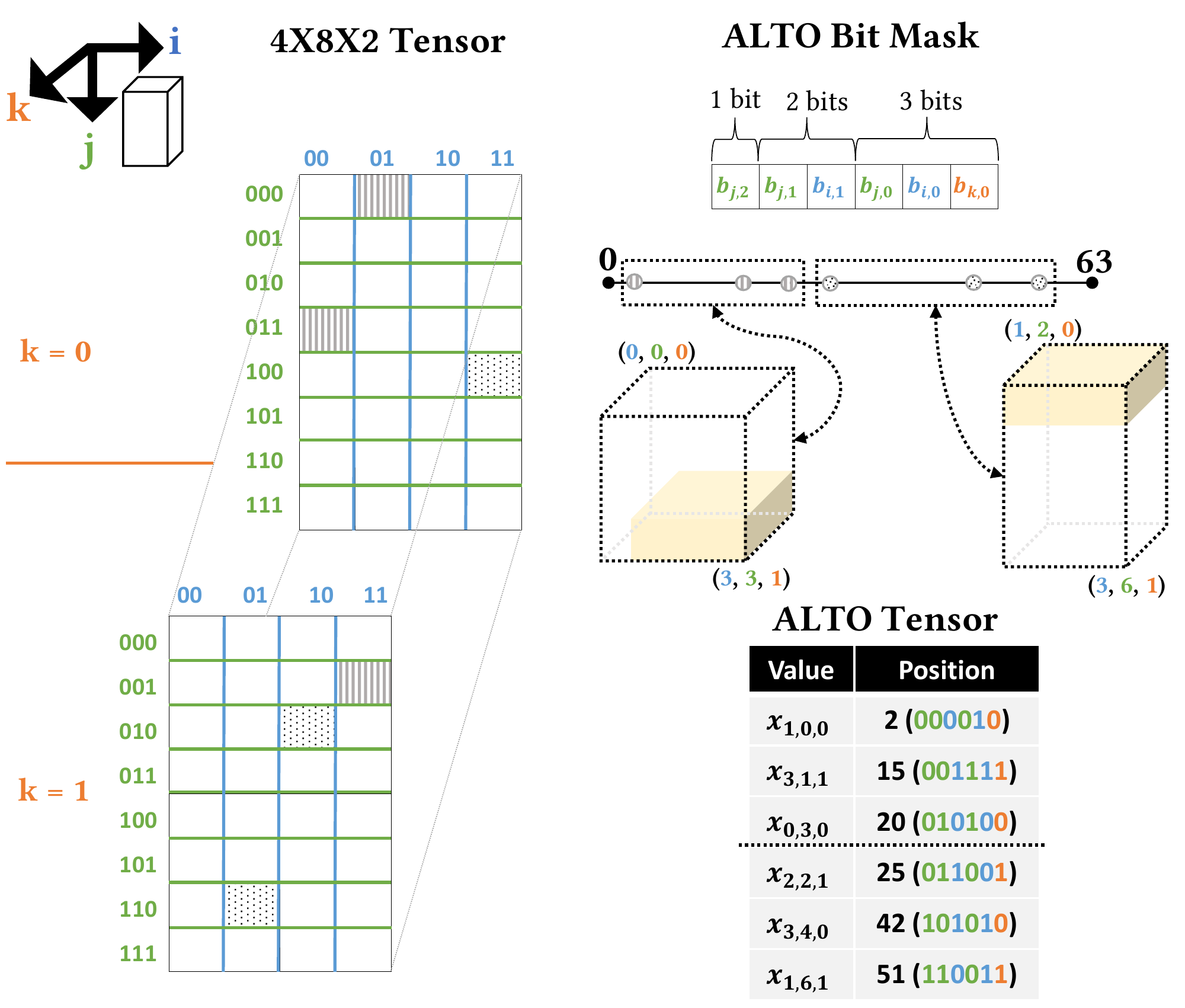}
	\vspace*{-17pt}
	\caption{\ALTO partitioning of the example in Figure~\ref{fig:alto_encoding},
	which generates balanced partitions	in terms of workload (nonzero elements) for efficient parallel execution.
	}
\label{fig:alto_parallel}
\end{figure}

To decouple the performance of sparse tensor computations from the distribution of nonzero elements, \ALTO eliminates the workload imbalance and generates perfectly balanced partitions. 
Figure~\ref{fig:alto_parallel} depicts an example of \ALTO's workload decomposition when applied to the sparse tensor in Figure~\ref{fig:alto_encoding}.
Moreover, \ALTO divides the line segment containing the nonzero elements of the target tensor into smaller line segments, all of which have the same number of nonzeros, thus perfectly splitting the workload. Therefore, in Figure~\ref{fig:alto_parallel}, \ALTO partitions the linearized tensor into two line segments: $[2-20]$ and $[25-51]$. Although the resulting line segments have different lengths (i.e., 18 and 26), they have the same number of nonzeros elements.

Once the linearized sparse tensor is divided into multiple line segments, \ALTO identifies the basis mode intervals (coordinate ranges) of the multi-dimensional subspaces that correspond to these segments. For example, the line segments $[2-20]$ and $[25-51]$ correspond to three-dimensional subspaces bounded by the mode intervals $\{[0-3], [0-3], [0-1]\}$ and $\{[1-3], [2-6], [0-1]\}$, respectively. While the derived multi-dimensional subspaces of the line segments may overlap, as highlighted in yellow in Figure~\ref{fig:alto_parallel}, each nonzero element is assigned to exactly one line segment. That is, \ALTO imposes a partitioning on a given linearized tensor that generates a disjoint set of \textit{non-overlapping and balanced line segments}, yet it does not guarantee that such a partitioning will decompose the multi-dimensional space of the tensor into non-overlapping subspaces. In contrast, the prior sparse tensor 
formats decompose the multi-dimensional space into non-overlapping (yet highly imbalanced) regions, namely, tensor slices and fibers in CSF-based formats and multi-dimensional spatial blocks in block-based formats (e.g., HiCOO). 

More formally, a set of $L$ line segments partitions a linearized \ALTO tensor \TENSOR{X}, which encodes a mode-$N$ sparse tensor, such that \(\TENSORI{X} = \TENSORI{X}_1 \cup \TENSORI{X}_2 \cdots \cup \TENSORI{X}_L\) and
\(\TENSORI{X}_{i} \cap \TENSORI{X}_{j} = \phi \forall i~\text{and}~j\), where $i \neq j$.
Each line segment \TENSOR{X}$_{i}$ is an ordered set of nonzero elements that are bounded in an $N$-dimensional space by a set of $N$ closed mode intervals
\(T_{i}~=~\{ [T_{i,1}^{s}, T_{i,1}^{e}], [T_{i,2}^{s}, T_{i,2}^{e}], \cdots [T_{i,N}^{s}, T_{i,N}^{e}]\}\),
where each mode interval $T_{i,j}$ is delineated by a start $T_{i,j}^{s}$ and an end $T_{i,j}^{e}$. The intersection of two sets of mode intervals represents the subspace overlap between their corresponding line segments (partitions).   

\subsection{Adaptive Tensor Operations}\label{sec:approach-ops}

Since high-dimensional data analytics is becoming increasingly popular in rapidly evolving areas~\cite{ho2014marble, papalexakis2016tensors, sidiropoulos2017tensor, kobayashi2018extracting}, a fundamental goal of the proposed \ALTO format is to deliver superior performance without compromising the productivity of end users to allow fast development of tensor algorithms and operations.
Algorithm~\ref{fig:mttkrp-algo} illustrates the popular MTTKRP tensor operation using the \ALTO format. First, unlike CSF-based formats, \ALTO enables end users to perform tensor operations using \textit{a unified code implementation that works on a single copy of the sparse tensor}, regardless of the different mode orientations of such operations. Second, by decoupling the representation of a sparse tensor from the distribution of its nonzero elements, \ALTO does not require \textit{manual} tuning to select the optimal \textit{format parameters} for this tensor, in contrast to block-based storage approaches such as HiCOO. Instead, the \ALTO format is automatically generated based on the shape and dimensions of the target sparse tensor (as explained in $\S$\ref{sec:approach-gen}).  

\begin{algorithm}[htb]
\begin{algorithmic}[1]
\Require A third-order \ALTO sparse tensor \TENSOR{X} $\in$ \REALIJK with $M$ nonzero elements, dense factor matrices \MATRIX{A} $\in$ \REALTWO{I}{R}, \MATRIX{B} $\in$ \REALTWO{J}{R}, and \MATRIX{C} $\in$ \REALTWO{K}{R} 
\Ensure Updated dense factor matrix \MATRIX{\~{A}} $\in$ \REALTWO{I}{R}
\For{ $x = 1, \dots, M$}
\State $i = \mathbb{EXTRACT}(pos(x), MASK(1))$ \Comment{De-linearization.}
\State $j = \mathbb{EXTRACT}(pos(x), MASK(2))$
\State $k = \mathbb{EXTRACT}(pos(x), MASK(3))$
    \For{ $r = 1, \dots, R$}
    \State \MATRIX{\~{A}}$(i,r)$ $+= val(x) \times$ \MATRIX{B}$(j,r) \times$ \MATRIX{C}$(k,r)$
    \EndFor
\EndFor
\Return \MATRIX{\~{A}}
\end{algorithmic}
\caption{Mode-$1$ sequential MTTKRP-\ALTO algorithm.}
\label{fig:mttkrp-algo}
\end{algorithm}

\begin{algorithm}[htb]
\newcommand{\algcolor}[2]{\hspace*{-\fboxsep}\colorbox{#1}{\parbox{\dimexpr\linewidth-\fboxsep}{#2}}}
\newcommand{\algemph}[1]{\algcolor{lighter-gray}{#1}}
\begin{algorithmic}[1]
\Require A third-order \ALTO sparse tensor \TENSOR{X} $\in$ \REALIJK with $M$ nonzero elements, dense factor matrices \MATRIX{A} $\in$ \REALTWO{I}{R}, \MATRIX{B} $\in$ \REALTWO{J}{R}, and \MATRIX{C} $\in$ \REALTWO{K}{R} 
\Ensure Updated dense factor matrix \MATRIX{\~{A}} $\in$ \REALTWO{I}{R}
\For{ $l = 1, \dots, L$ \textbf{in parallel}} \Comment{\ALTO line segments.}
    \For{ $\forall x \in$ \TENSOR{X}$_{l}$}
    \State $i = \mathbb{EXTRACT}(pos(x), MASK(1))$ \Comment{De-linearization.}
    \State $j = \mathbb{EXTRACT}(pos(x), MASK(2))$
    \State $k = \mathbb{EXTRACT}(pos(x), MASK(3))$
        \For{ $r = 1, \dots, R$}
        \State \colorbox{lighter-gray}{\MATRIX{Temp}$_{l}(i-T_{l,1}^{s},r)$ $+= val(x) \times$ \MATRIX{B}$(j,r) \times$ \MATRIX{C}$(k,r)$}\vspace*{-\fboxsep}
        \State \colorbox{light-gray}{$\mathbb{ATOMIC}($\MATRIX{\~{A}}$(i,r)$ $+= val(x) \times$ \MATRIX{B}$(j,r) \times$ \MATRIX{C}$(k,r)$)}
        \EndFor
    \EndFor
\EndFor
\item[]
\algemph{
\For{ $b = 1, \dots, I$ \textbf{in parallel}} \Comment{Pull-based accumulation.}
    \For{ $\forall l$  where $b \in [T_{l,1}^{s}, T_{l,1}^{e}]$}
        \For{ $r = 1, \dots, R$}
        \State \MATRIX{\~{A}}$(b,r)$ $+=$ \MATRIX{Temp}$_{l}(b-T_{l,1}^{s},r)$
        \EndFor
    \EndFor
\EndFor
}
\Return \MATRIX{\~{A}}
\end{algorithmic}
\caption{Adaptive parallel execution of mode-$1$ MTTKRP-\ALTO kernel. \ALTO automatically uses  \colorbox{lighter-gray}{local storage} or \colorbox{light-gray}{atomics}, based on the reuse of output fibers, to resolve the update conflicts. }
\label{fig:mttkrp-par-algo}
\end{algorithm}

Because processing the nonzero elements in parallel (line~1 in Algorithm~\ref{fig:mttkrp-algo}) can result in write conflicts across threads (line~6 in Algorithm~\ref{fig:mttkrp-algo}), we devise an effective parallel execution and synchronization algorithm that handles these conflicts by exploiting the inherent data reuse of target tensors.
Algorithm~\ref{fig:mttkrp-par-algo} describes the proposed workload distribution and scheduling scheme using a representative parallel MTTKRP operation that works on a sparse tensor stored in the \ALTO format. After \ALTO imposes a partitioning on a given sparse tensor, as detailed in $\S$\ref{sec:approach-par}, each partition can be assigned to a different thread. To resolve the update/write conflicts that may happen during parallel sparse tensor computations,
\ALTO uses \textit{an adaptive conflict resolution} approach that \textit{automatically} selects the appropriate global synchronization technique (highlighted by the different gray backgrounds) across threads based on the reuse of the target fibers. This metric represents the average number of nonzero elements per fiber (the generalization of a matrix row/column) and it is estimated by simply dividing the total number of nonzero elements by the number of fibers along the target mode. When a sparse tensor operation suffers from limited fiber reuse, \ALTO resolves the conflicting updates across its line segments (partitions) using direct atomic operations (line~8). Otherwise, it uses a limited amount of temporary (local) storage to keep the local updates of different partitions (line~7) and then merges the conflicting global updates (lines~12--18) using an efficient pull-based parallel reduction, where the final results are computed by pulling the partial results from the \ALTO partitions. 

\ALTO considers the fiber reuse large enough to use local staging memory for conflict resolution, if the average number of nonzero elements per fiber is \textit{more than} the maximum cost of using this two-stage (buffered) accumulation process, which consists of initialization (omitted for brevity), local accumulation (line~7 in Algorithm~\ref{fig:mttkrp-par-algo}), and global accumulation (lines~12--18). Specifically, the buffered accumulation cost is four memory operations (two read and two write operations) at most, i.e., in the worst (no reuse) case.
As explained in $\S$\ref{sec:approach-par}, each line segment \TENSOR{X}$_{l}$ is bounded in an $N$-dimensional space by a set of $N$ closed mode intervals $T_{l}$, which is computed during the partitioning of an \ALTO tensor; thus, the size of the temporary storage accessed during the accumulation of \TENSOR{X}$_{l}$'s updates along a mode $n$ is directly determined by the mode interval $[T_{l,n}^{s}, T_{l,n}^{e}]$ (see lines~7 and 13). 

Finally, \ALTO allows automated generation and use of helper flags to further reduce the conflict resolution overhead in the sparse tensors that suffer from limited fiber reuse. That is, it exploits the unused (do-not-care) bits in the linearized index to encode if a nonzero element is a boundary or internal (exclusive) element along a mode with limited fiber reuse. Based on this information, \ALTO determines whether to execute the global update (line~8 in Algorithm~\ref{fig:mttkrp-par-algo}) as an atomic operation (for boundary elements) or a normal write (for internal elements).

\section{Evaluation}\label{sec:perf}
We evaluate \ALTO against the state-of-the-art sparse tensor representations in terms of tensor storage, parallel performance and execution time, and format generation cost. We conduct a thorough study of key tensor decomposition operations ($\S$\ref{sec:cpd-als}) and demonstrate the performance characteristics of \ALTO not only compared to the prior formats, but also relative to an oracle that selects the \textit{best} mode-agnostic and mode-specific format for each tensor dataset.

\subsection{Implementation}\label{sec:impl}
We implemented \ALTO as a C++ library and used OpenMP~\cite{dagum1998openmp} for multi-threaded execution. The implementation utilizes automatic vectorization and loop unrolling optimizations~\cite{bik2002automatic}, and it uses templates~\cite{vandevoorde2002} for generalized support of tensors with arbitrary sizes. Specifically, we use a generic type to represent our mode-agnostic indexing metadata, which allows the same code to support linearized indices of different widths.
This template-based approach avoids code duplication and reduces the time and effort required to port \ALTO to other hardware platforms.    

To improve the performance of sparse tensor kernels, the state-of-the-art tensor libraries specialize these kernels for different tensor orders (e.g., 3D and 4D tensors). In our library, a canonical tensor operation, such as MTTKRP, has an entry point that acts as a dispatcher, which invokes the generic implementation or more specialized versions of this implementation when available. This malleable approach leverages the compiler to transparently generate optimized code for common tensor orders and/or for typical decomposition ranks (called rank specialization). However, for a fair comparison with existing tensor libraries, we report the performance of \ALTO \textit{without rank specialization} and discuss the potential performance improvement of such an optimization.  

We also incorporated ALTO into the popular SPLATT library~\cite{Smith2015} (i.e., replaced the CSF-based MTTKRP operation with the ALTO-based MTTKRP) to validate its usability in the CPD-ALS algorithm.
Given the same initial values, our implementation calculates identical factor matrices as the original SPLATT implementation and shows the same convergence properties (i.e., same number of iterations to convergence and fit to the original tensor).
Our implementation also shows similar fit compared to the Tensor Toolbox~\cite{Bader2007} CPD-ALS implementation.

\subsection{Experimental Setup}
\subsubsection{Platform}
All experiments were conducted on an Intel Xeon Platinum 8280 CPU with \ac{CLX} microarchitecture. It consists of two sockets, each with 28 physical cores running at a fixed frequency of 1.8\,\GHZ ~for accurate measurements. The server has 384 GiB of memory and it runs CentOS~7.7 Linux distribution. 
The code is built using Intel C/C++ compiler (version~19.1.3) with the optimization flags \texttt{-O3} \texttt{-xCORE-AVX512} \texttt{-qopt-zmm-usage=high} to fully utilize vector units. Unless otherwise stated, the parallel experiments use all hardware threads ($112$) on the target platform. We report the performance numbers as an average over 100 iterations/runs, after a warmup iteration as per prior tensor libraries.
For performance counter measurements and thread pinning, we use the LIKWID tool suite~v5.1.0~\cite{likwid}.

\subsubsection{Datasets}
For a comprehensive evaluation, the experiments consider a gamut of real-world tensor datasets with various characteristics. These tensors are often used in related works and they are publicly available in the FROSTT~\cite{frostt} and HaTen2~\cite{haten2_ICDE2015} repositories. Table~\ref{tab:problems} shows the detailed features of the target datasets, ordered by size, in terms of dimensions, number of nonzero elements (\#NNZs), and density. To make the results clear and interpretable, the tensors are classified based on the average reuse of their fibers into high, medium, or limited reuse. We consider the fibers along a given mode to have high reuse, if they are reused more than eight times on average; when the fibers are reused between five to eight times, they have medium reuse; otherwise, the fibers suffer from limited reuse. Since the target tensor operations access fibers along all modes, a tensor with one or more modes of limited/medium reuse is considered to have an overall limited/medium fiber reuse.    

\begin{table}[tb]
\small  
\centering
\caption{Characteristics of the target sparse tensors.}
\vspace*{-5pt}
\label{tab:problems}
\begin{tabular}{|p{1.1cm}|p{2.43cm}|p{0.83cm}|p{1.18cm}|p{1.2cm}|}
\hline
\textbf{Tensor} & \textbf{Dimensions} & \textbf{\#NNZs} & \textbf{Density} & \textbf{Fib. reuse}\\\toprule\hline
\textsc{lbnl} & $1.6K\times4.2K\times1.6K\times4.2K\times868.1K$ & $1.7M$ & $4.2\times10^{-14}$ & Limited \\\hline
\textsc{nips} & $ 2.5K\times 2.9K\times 14K\times17$ & $3.1M$ & $1.8\times10^{-06}$& High\\\hline 
\textsc{uber} & $ 183\times24\times 1.1K\times1.7K$&  $3.3M$ & $3.8\times10^{-04}$& High\\\hline 
\textsc{chicago} & $ 6.2K\times24\times77\times32$& $5.3M$& $1.5\times10^{-02}$& High\\\hline 
\textsc{vast} & $165.4K\times11.4K\times2\times100\times89$& $ 26M$& $7.8\times10^{-07}$& High\\\hline 
\textsc{darpa} & $22.5K\times22.5K\times23.8M$& $28.4M$& $2.4\times10^{-09}$& Limited \\\hline 
\textsc{enron} & $ 6K\times5.7K\times244.3K\times1.2K$& $54.2M$& $5.5\times10^{-09}$& High \\\hline 
\textsc{nell-2} & $12.1K\times9.2K\times28.8K$& $76.9M$& $2.4\times10^{-05}$& High\\\hline 
\textsc{fb-m} & $23.3M\times23.3M\times166$ & $99.6M$& $1.1\times10^{-09}$& Limited\\\hline 
\textsc{flickr} & $319.7K\times28.2M\times1.6M\times731$& $112.9M$& $1.1\times10^{-14}$& Limited\\\hline 
\textsc{deli} & $532.9K\times17.3M\times2.5M\times1.4K$& $140.1M$& $4.3\times10^{-15}$& Medium\\\hline 
\textsc{nell-1} & $2.9M\times2.1M\times25.5M$& $143.6M$& $9.1\times10^{-13}$& Medium\\\hline 
\textsc{amazon} & $4.8M\times1.8M\times1.8M$& $1.7B$ & $1.1\times10^{-10}$& High\\\hline 
\textsc{patents} & $46\times239.2K\times239.2K$& $3.6B$ & $1.4\times10^{-03}$& High\\\hline 
\textsc{reddit} & $8.2M\times177K\times8.1M$& $4.7B$& $4.0\times10^{-10}$& High\\\hline 
\end{tabular}
\vspace*{-2pt}
\end{table}

\subsubsection{Configurations}
We evaluate the proposed \ALTO format compared to the mode-agnostic COO and HiCOO formats~\cite{li2018hicoo} as well as the mode-specific CSF formats~\cite{Smith2015a, Smith2015}. Specifically, we use the latest code
of the state-of-the-art sparse tensor libraries for CPUs, namely, ParTI\footnote{Available at: \url{https://github.com/hpcgarage/ParTI}} and SPLATT.\footnote{Available at: \url{https://github.com/ShadenSmith/splatt}} 
We report the best-achieved performance across the different configurations of the COO format; that is, with or without thread privatization (which keeps local copies of the output factor matrix). For the HiCOO format, its performance and storage are highly sensitive to the block and superblock (SB) sizes, which benefit from tuning. Since the current HiCOO implementation does not auto-tune the format parameters or consider the required tuning time in the format generation cost, we use a block size of 128 ($2^{7}$) and two superblock sizes of $2^{10}$ and $2^{14}$ according to prior work~\cite{sun2020sptfs}, and we report the performance of each format variant 
(``HiCOO-SB10'' and ``HiCOO-SB14'').
We evaluate two variants of the mode-specific formats: CSF and CSF with tensor tilling (``CSF-tile''), both of which use $N$ representations (``SPLATT-ALL'') for an order-$N$ sparse tensor to achieve the best performance.\footnote{In \textsc{reddit} dataset, SPLATT runs out of memory. While keeping fewer tensor copies is possible, it significantly degrades the performance on non-root modes due to using different recursion- and lock-based algorithmic variants.} Similar to previous studies~\cite{Smith2015a, Smith2015, choi2018blocking}, the experiments use double-precision arithmetic and $64$-bit (native word) integers. While the target datasets require a linearized index of size between $32$ and $80$ bits, we configured \ALTO to select the size of its linearized index to be multiples of the native word size (i.e., $64$ and $128$ bits) for simplicity. We use a decomposition rank $R=16$ for all experiments.

\subsection{Summary of Performance Metrics}
\begin{figure}[tb]
\centering
\includegraphics[trim=55 275 55 285, clip, width=1.0\linewidth]{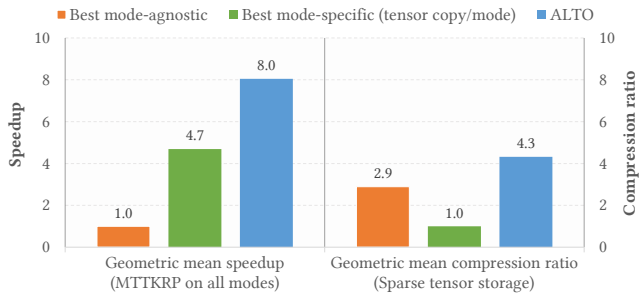}
\vspace*{-15pt}
	\caption{
	    The performance metrics of the \ALTO format (higher is better) in comparison with an oracle that selects the best mode-agnostic or mode-specific format variant for tensor datasets.} 
	\vspace*{-4pt}
\label{fig:perf-sum}
\end{figure}

Figure~\ref{fig:perf-sum} compares the performance characteristics of the proposed \ALTO format to an oracle that selects the best format for target tensors. The oracle considers two distinct types of sparse tensor formats: 1) mode-agnostic or general formats (COO, HiCOO-SB10, and HiCOO-SB14), which use a single tensor representation, and 2) mode-specific formats (CSF and CSF-tile), which keep multiple tensor copies (one per mode) for best performance. The results show that \ALTO outperforms the best mode-agnostic as well as mode-specific formats in terms of the speedup of tensor operations (MTTKRP on all modes) and the tensor storage. Specifically, \ALTO achieves a geometric mean speedup of $8\times$ compared to the best general formats, while delivering a geometric mean compression ratio of $4.3\times$ relative to the best mode-specific formats. What follows is a detailed analysis and discussion of the performance results.  

\begin{figure*}[htb]
\centering
\includegraphics[trim=53 335 55 340, clip, width=0.98\linewidth]{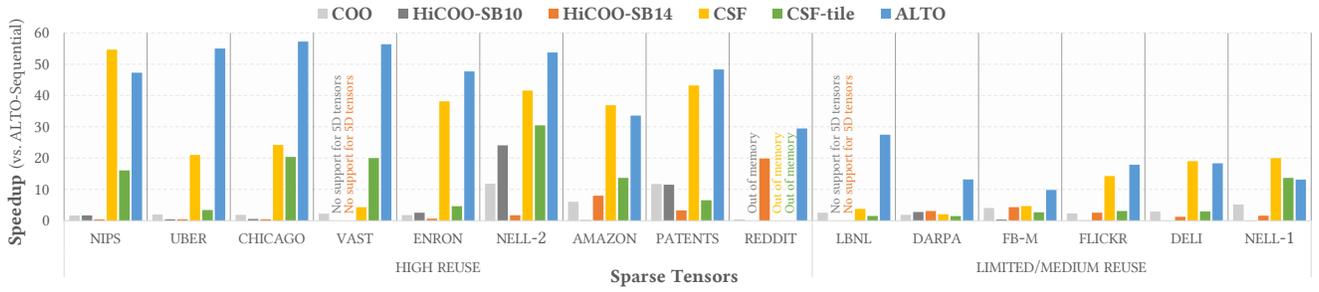}
\vspace*{-12pt}
	\caption{The overall parallel performance/speedup of MTTKRP (all modes) using the different sparse tensor formats. The speedup is reported compared to the optimized sequential MTTKRP-ALTO version to show the parallel scalability. The sparse tensors are categorized and then sorted in increasing order of their size (number of nonzero elements).} 
\label{fig:scaling}
\end{figure*}

\begin{figure*}[htb]
\centering
\vspace*{-5pt}
\includegraphics[trim=55 340 55 340, clip, width=0.98\linewidth]{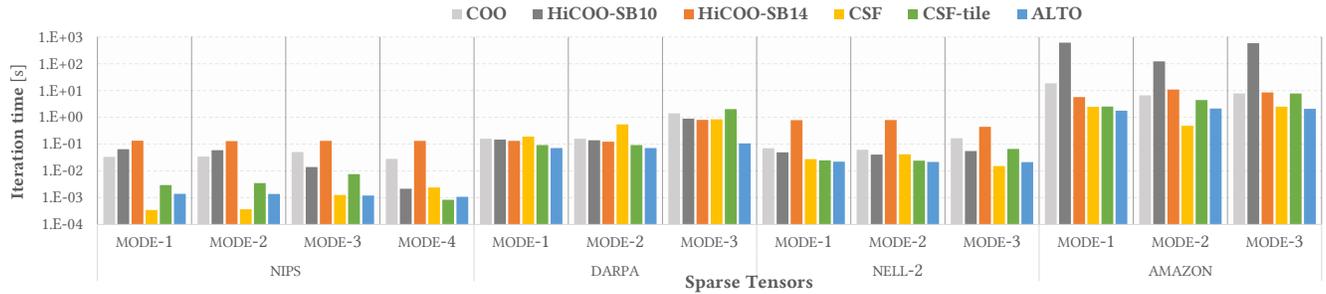}
\vspace*{-10pt}
	\caption{The execution time of parallel MTTKRP implementations across different modes for select tensors with various characteristics.}
	\vspace*{-8pt}
\label{fig:perf-mode}
\end{figure*} 
  
\subsection{Performance Results and Analysis}
To evaluate the parallel performance and scalability of the \ALTO format, Figure~\ref{fig:scaling} shows the speedup of the different parallel implementations of MTTKRP, which dominates the execution time of tensor decomposition methods.
Unlike prior formats, the parallel performance of \ALTO depends on the inherent data reuse of sparse tensors rather than the spatial distribution of their nonzero elements. 
As explained in Algorithm~\ref{fig:mttkrp-par-algo}, the parallel MTTKRP-\ALTO algorithm consists of two stages: 1) computing the output fibers of the target mode using the input fibers along all other modes, and 2) merging the conflicting updates of output fibers across threads. 
\ALTO demonstrates linear scaling for the sparse tensors with high reuse (on the left of the figure). Specifically, it achieves a geometric mean speedup of $47\times$ on $56$ cores (i.e., more than $80\%$ parallel efficiency) by exploiting the data locality of input fibers and by locally computing the partial updates of output fibers in higher levels of the memory system hierarchy. This way, the overhead of merging these partial updates is amortized over the large number of output fiber reuse. The analysis of performance counters, as detailed below, shows that most of the input/output fiber data is accessed in the cache, which strongly reduces the pressure on the main memory.
For the \textsc{amazon}\- and \textsc{reddit} tensors, their extreme sparsity impacts the parallel scalability compared to other high-reuse cases, due to the reduction of useful work performed over the application's memory footprint. In addition, when sparse tensors have limited/medium data reuse, as shown on the right of Figure~\ref{fig:scaling}, \ALTO has a sub-linear scaling (a geometric mean speedup of $16\times$ on 56 cores) due to the large number of accesses to main memory. On our test platform,
the STREAM~Triad~\cite{McCalpin1995} bandwidth is \(10\)\,\GBS\ and \(210\)\,\GBS\ for a single core and a full node of 56 cores, respectively. Hence, \ALTO obtains around 76\% of the maximum realizable speedup (\(21\times\)) in such memory-bound cases.

On the contrary, the performance of prior sparse tensor formats varies widely due to their sensitivity to the irregular shapes and data distributions of higher-order tensors. As a result, the location-based formats (HiCOO-SB10/SB14, CSF, and CSF-tile) suffer from workload imbalance and ineffective compression depending on the distribution of the nonzero elements of each dataset. Additionally, based on the update conflicts across blocks/superblocks, the parallel schedule of block-based formats (HiCOO-SB10/SB14) can encounter serialization and/or substantial parallelization overhead. However, in a few cases, namely, \textsc{nips}, \textsc{amazon}, and \textsc{nell-1} tensors, the mode-specific CSF format shows slightly better performance ($1.2\times$ geometric-mean speedup) than the mode-agnostic \ALTO format by keeping multiple sparse tensor representations along different mode orientations, which in turn significantly increases the storage of CSF (by $2.5\times$--$4.5\times$ compared to \ALTO, as shown in Figure~\ref{fig:mem}). 

\subsubsection{Performance across Modes}\label{ssec:mode-behav}
Figure~\ref{fig:perf-mode} demonstrates the runtime variations of the different formats while performing MTTKRP across tensor modes. The selected sparse tensors exhibit different characteristics in terms of shape, size, density, and data reuse. Compared to the other formats, \ALTO has a relatively consistent performance regardless of the mode orientation of tensor operations. In the \textsc{darpa} tensor, the execution time of \ALTO along mode-3 is higher than mode-1 and mode-2 due to the limited fiber reuse of mode-3. While fibers along all modes are read during MTTKRP\- execution, only the output fibers along the target mode are updated. Since the memory read bandwidth is typically higher than the write bandwidth, the limited fiber reuse of mode-3 has more impact on mode-3 MTTKRP compared to other modes. In contrast, the other sparse formats suffer from significant performance variations across modes (note the use of logarithmic scale). While the execution time of the mode-specific CSF formats is expected to change based on the characteristics of the target mode, as they use a distinct tensor representation for each mode orientation, the results show that the parallel performance of block-based formats varies widely due to the different update conflicts and parallelism degrees of tensor blocks/superblocks across modes.          

\subsubsection{Roofline Analysis}
To better understand the performance characteristics of MTTKRP using the \ALTO format, we created a \Rlm~\cite{roofline:2009} for the test platform and collected performance counters across a set of representative parallel runs.
For the \Rlm, an upper performance limit \(\Pi\) is given by \(\Pi = \mathrm{min}(\Pi_{\mathrm{peak}}, B_{\mathrm{peak}} \times OI)\), where
\(\Pi_{\mathrm{peak}}\) is the theoretical peak performance, \(B_{\mathrm{peak}}\) is the peak memory bandwidth, and \(OI\) is the \emph{operational intensity} (i.e., the ratio of executed floating-point operations per byte).
In addition, we enhance our \Rlm ~to consider the bandwidth of the different cache levels. While the L2 cache, L3 cache, and main memory bandwidth in our model are phenomenological, i.e., measured using likwid-bench, L1 bandwidth measurements are error-prone. Therefore, we use the theoretical L1 load bandwidth of two cache lines per cycle per core for this particular \rl.
The peak performance, \(\Pi_{\mathrm{peak}}\), is calculated based on the ability of the cores to execute two fused multiply-add~(FMA) instructions on eight-element double precision vector registers~(due to the availability of AVX-512) per cycle.

\begin{figure}[tb]
\centering
\includegraphics[width=0.9\linewidth]{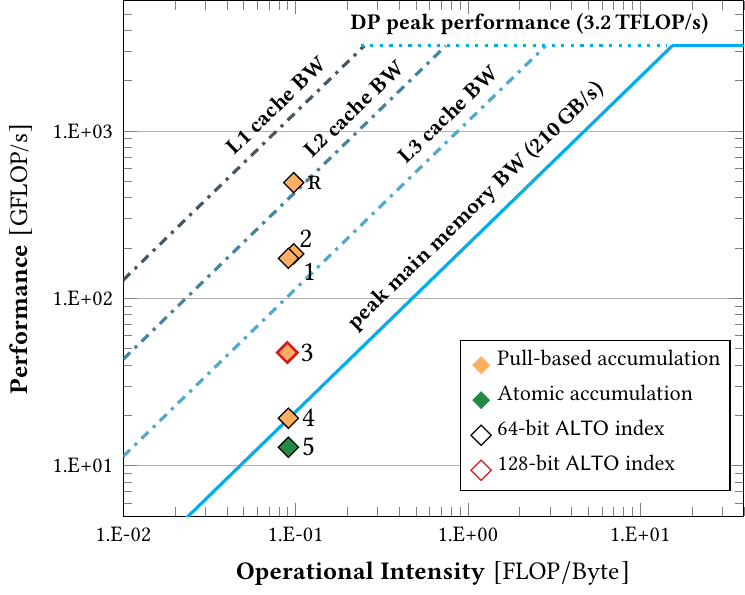}
\centering
\vspace*{-10pt}
\caption{The parallel performance of \ALTO across MTTKRP runs: \textsc{nips} mode-4 without~\protect\bcircled{1} and with~\protect\bcircled{\textsc{r}} rank specialization, \textsc{nell-2} mode-3~\protect\bcircled{2}, \textsc{amazon} mode-1~\protect\bcircled{3}, \textsc{darpa} mode-1~\protect\bcircled{4} and mode-3~\protect\bcircled{5}.} 
\vspace*{-10pt}
\label{fig:roofline}
\end{figure}

Figure~\ref{fig:roofline} shows the performance of the parallel MTTKRP-\ALTO computations (lines 1-11 in Algorithm~\ref{fig:mttkrp-par-algo}) for several representative cases.
While the memory-intensive MTTKRP operation suffers from low operational intensity, \ALTO can still exceed the peak main memory bandwidth by exploiting the inherent data reuse and by efficiently resolving the update conflicts. 
The \textsc{nips} tensor~\bcircled{1} is relatively small and it has high reuse of output fibers, which can be held locally in L1/L2 caches. The performance counters show a superior cache hit rate while computing output fibers, which explains the high effective bandwidth and floating-point throughput.
Mode-3 of \textsc{nell-2}~\bcircled{2} has a smaller amount of fiber reuse and it accesses a larger number of local output fibers compared to \textsc{nips}~\bcircled{1}. 
Nonetheless, \ALTO efficiently utilizes all threads to do useful work and amortizes the parallelization overheads due to the larger number of nonzero elements, and we observe a slightly higher performance than~\bcircled{1}. 
While the \textsc{amazon} tensor~\bcircled{3} has high fiber reuse, its extreme sparsity and large size increase the memory pressure.
In addition, although the tensor can be encoded using a $72$-bit linearized index, our current \ALTO implementation uses a $128$-bit linearized index (twice the native word size) for simplicity. Together these factors increase the data volume and the de-linearization overhead, which in turn reduces the effective floating-point throughput compared to other tensors with high data reuse~\bcircled{1} and~\bcircled{2}.

In contrast, the \textsc{darpa} tensor has limited fiber reuse (along mode-3) which restricts the performance compared to the previous cases. During the computation of MTTKRP on mode-3~\bcircled{5}, the limited reuse of output fibers does not justify the use of local/temporary memory to resolve the conflicting updates. As a result, \ALTO automatically chooses to use atomic updates for conflict resolution. While \textsc{darpa} mode-1~\bcircled{4}  has higher reuse of output fibers, 
comparable to~\bcircled{2}, reading the input fibers along mode-3 dominates the execution time and results in more data traffic across the memory system levels because of the limited temporal locality; thus, we can observe that its performance is bounded by the memory bandwidth.  

\subsection{Impact of Rank Specialization}
The generic (template-based) implementation of \ALTO (which is discussed in $\S$\ref{sec:impl}) not only allows effortless specialization of sparse tensor kernels for common tensor orders (e.g., 3D and 4D tensors), but also makes it possible to specialize these kernels for typical ranks --- an optimization that we call rank specialization.
For a fair comparison with the prior sparse tensor implementations, all other performance results in this paper using \ALTO \textit{do not leverage rank specialization}. Here, we show the potential of this feature to further improve the performance by allowing the compiler to gain more insight into the length of rank-wise update operations.
This way, the compiler can generate more optimized code, which leads to a better overall performance, as shown for \textsc{nips} mode-4~\bcircled{\textsc{r}}.
The analysis of the generated assembly files with OSACA~\cite{osaca} shows that rank specialization results in better control structures with fewer branches compared to the standard implementation.
Additionally, the compiler can reduce the pressure on the load units and, thus, decreases the execution time by more than $2\times$ in \textsc{nips}~\bcircled{\textsc{r}}. 

\begin{figure}[tb]
\centering
\includegraphics[trim=60 270 60 270, clip, width=1.0\linewidth]{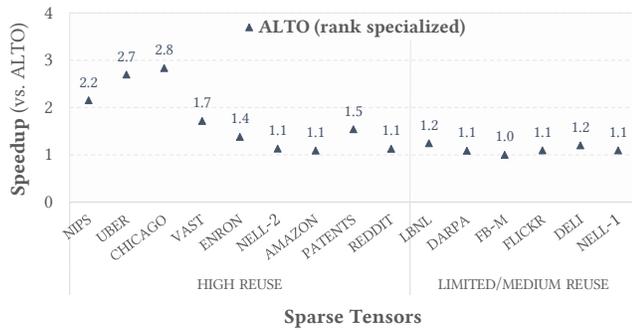}
\vspace*{-16pt}
	\caption{The speedup of \ALTO using rank specialization relative to the default \ALTO parallel implementation. The sparse tensors are categorized and then sorted in an increasing order of their size.}
	\vspace*{-8pt}
\label{fig:perf-rs}
\end{figure}

\begin{figure*}[htb]
\centering
\includegraphics[trim=45 343 55 340, clip, width=0.98\linewidth]{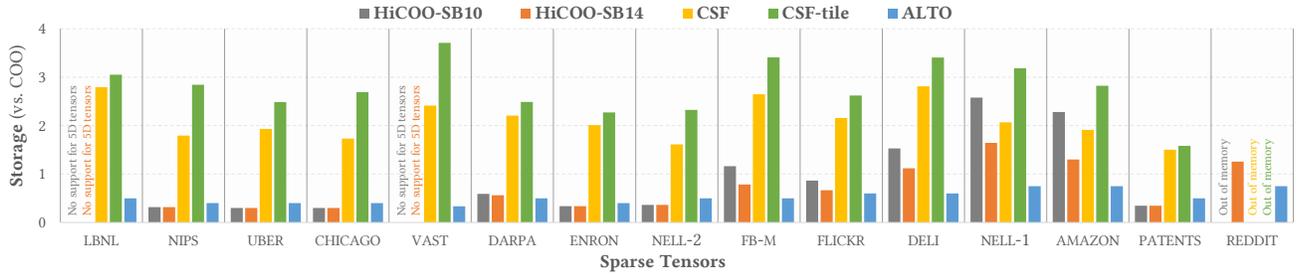}
\vspace*{-10pt}
	\caption{The sparse tensor storage using the different formats compared to COO. The tensors are sorted in an increasing order of their size.}
\label{fig:mem}
\end{figure*}

\begin{figure*}[htb]
\centering
\vspace*{-8pt}
\includegraphics[trim=45 340 55 340, clip, width=0.98\linewidth]{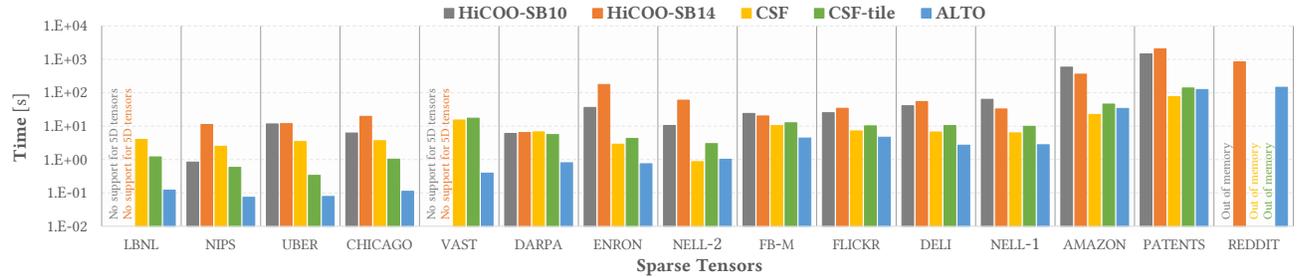}
\vspace*{-12pt}
	\caption{The format construction cost in seconds. The sparse tensors are sorted in an increasing order of their size.}
	\vspace*{-5pt}
\label{fig:perf-format}
\end{figure*}

Figure~\ref{fig:perf-rs} shows a comprehensive comparison between the rank specialized \ALTO version and the default parallel version used in prior experiments.
For smaller tensors with high reuse, the compiler manages to optimize the computations and to improve the performance by more than $2\times$.
In larger tensors as well as tensors with limited reuse, we can still observe a performance benefit of more than 10\%, leading to an overall geometric mean speedup of \(1.4\times\) across the different sparse tensors.

\subsection{Memory Storage}
Figure~\ref{fig:mem} details the relative storage of the sparse tensors using the proposed \ALTO format as well as the block-based (HiCOO-SB10/SB14) and tree-based (CSF and CSF-tile) formats. Compared to the de facto COO format, \ALTO always requires less storage space due to its efficient linearization of the multi-dimensional space, as explained in $\S$\ref{sec:approach-gen}. The mode-specific (CSF and CSF-tile) formats consume more storage space than COO because they keep multiple sparse representations for the different mode orientations. Overall, imposing a tilling over the tensors (CSF-tile) ends up increasing the storage requirements of CSF. Conversely, the memory consumption of the block-based formats is highly variable as it depends on the block/superblock sizes and the distribution of the nonzero elements in the multi-dimensional space. Thus, HiCOO can reduce the storage compared to COO when the number of nonzero elements per block is relatively high. However, as the sparsity of the tensors increases, the block-based formats consume more space, such as in the \textsc{deli}, \textsc{nell-1}, \textsc{amazon}, and \textsc{reddit} tensors.

\subsection{Format Generation Cost}\label{sec:perf-format} 
Figure~\ref{fig:perf-format} compares the cost of constructing the different sparse tensor formats from a sparse tensor in the COO format. Since \ALTO works on a linearized rather than a multi-dimensional representation of tensors, the cost of sorting the nonzero elements in its linearized storage (which is the most expensive step in the format generation) is substantially reduced because of the decrease in the number of comparison operations. In addition to sorting the nonzero elements, the block-based HiCOO formats require costly clustering of these elements based on their multi-dimensional coordinates as well as scheduling of the resulting blocks/superblocks to avoid conflicts. As a result, \ALTO has more than an order-of-magnitude speedup relative to HiCOO in terms of the format conversion time. Although the tree-based CSF formats are generated from \textit{presorted} COO tensors, \ALTO still requires less time for format construction on average. In large-scale tensors, such as \textsc{amazon} and \textsc{patents}, \ALTO has comparable conversion cost to the tree-based formats as the tensor sorting time increases with the number of nonzero elements. Currently, the format generation of \ALTO leverages the C++ Standard Template Library (STL) and a further reduction of this construction cost is possible.

\section{Related Work and Discussion} \label{sec:related}
Due to the popularity of high-dimensional data analytics, a rich set of work has been recently developed to efficiently store sparse tensors and to optimize various tensor computations. Here, we discuss the related sparse tensor formats and operations as well as the conventional recursive data layouts.

\subsection{Sparse Tensor Formats}
Our study was motivated by the linearized coordinate (LCO) format~\cite{harrison2018high}, which also flattens sparse tensors but along a given orientation of tensor modes.
Hence, LCO requires either multiple tensor copies or on-the-fly permutation of tensors for efficient computation. 
In addition, the authors limit their focus to sequential algorithms, and it is not clear how LCO can be used to efficiently parallelize sparse tensor computations.

Recent extensions~\cite{liu2017unified, phipps2019software} to the list-based COO format reduce the synchronization overhead across threads using \textit{mode-specific} scheduling/permutation arrays. 
However, their output-oriented traversal of sparse tensors adversely affects the input data locality and/or result in random access of the nonzero elements~\cite{phipps2019software}. 
Moreover, storing this fine-grained scheduling information for all tensor modes can more than double the memory consumption, compared to the COO format~\cite{phipps2019software}. 

The block-based formats, such as HiCOO~\cite{li2018hicoo}, are highly sensitive to the characteristics of sparse tensors as well as the block size. When the blocking ratio (i.e., the number of blocks to the number of nonzero elements) is high, HiCOO can use more storage space than the simple COO format~\cite{li2018hicoo}. Due to the nonuniform (skewed) data distributions in the multi-dimensional space, the number of nonzero elements per block varies widely across HiCOO blocks, even after expensive mode-specific tensor permutations which in practice further increase workload imbalance~\cite{li2019efficient}. Moreover, such block-based formats, which use small data types for element indices, can end up under-utilizing the compute units and memory bandwidth because 1) mainstream architectures are word-oriented, which leads to limited efficiency of sub-word (narrow-width) 
operations~\cite{Abel19a, mittal2017survey}, and 2) applications need to generate large memory transactions to efficiently utilize the bandwidth of commodity DRAM chips~\cite{ghose2019demystifying}.

The mode-specific CSF format~\cite{Smith2015, Smith2015a} clusters the nonzero tensor elements into coarse-grained tensor slices and fibers, which limits its scalability on massively parallel architectures. To improve the performance on GPUs, recent CSF-based formats~\cite{nisa2019load, nisa2019efficient} expose more balanced and fine-grained parallelism but at the expense of substantial synchronization overheads and expensive preprocessing and format generation costs. The sparse tensor format selection (SpTFS) framework~\cite{sun2020sptfs} leverages deep learning models~\cite{zhao2018bridging, xie2019ia} to predict the best of COO, HiCOO, and CSF formats to perform the MTTKRP\- operation on a given sparse tensor. 

\subsection{Sparse Tensor Operations}
In addition to the widely used sparse tensor decomposition, which we extensively discussed in
previous sections, other important tensor algebra operations can benefit from the proposed \ALTO format. Examples of these operations include sparse tensor transposition and contraction as well as streaming tensor analysis. 

Sparse tensor transpositions are prevalent in high-dimensional data processing and analysis algorithms~\cite{mueller2020sparse}. The state-of-the-art approaches~\cite{wang2016parallel, mueller2020sparse} reduce these important operations to sorting the nonzero elements based on their multi-dimensional coordinates. \ALTO accelerates the tensor sorting process due to its linearized representation and recursive ordering of nonzero elements, which makes the sparse tensors stored in the \ALTO format amenable to partial radix-based sorting. 

Sparse tensor contractions emerge in many critical scientific domains~\cite{solomonik2014massively}, e.g., computational physics and quantum chemistry. These operations need random access into tensors, which can be supported by hash-based COO representations, such as Sparta~\cite{liu2021sparta}. \ALTO can further improve performance by using the target subset of encoding bits that represent the contract modes to quickly build and access a hash-based representation of linearized tensors.

To analyze infinite data sources, streaming tensor decomposition assembles batches of incoming data into a sequence of sparse sub-tensors and incrementally computes factor matrices~\cite{smith2018streaming, Soh:2021}. As shown in $\S$\ref{sec:perf-format}, \ALTO uses a fraction of the format generation time required by prior compressed formats, which makes it more suitable for accelerating such streaming algorithms. 

\vspace*{-5pt}
\subsection{Recursive Data Layouts}
Previous studies explored cache-oblivious~\cite{frigo1999cache, gustavson1997recursion} data layouts for optimizing memory-intensive operations, especially in the context of linear algebra. These data layouts exploit the underlying memory hierarchy without knowing its specific structure or configuration. Recursive algorithms and  storage schemes~\cite{frens1997auto, chatterjee2002recursive, elmroth2004recursive, peise2017algorithm} are popular in Basic Linear Algebra Subprogram~(BLAS) kernels, due to the regularity of such dense computations. 

Yzelman et al.~\cite{yzelman2009cache, yzelman2011two} proposed recursive data layouts for sparse matrices, based on hypergraph partitioning, to improve the performance of sparse-matrix, dense-vector (SpMV) multiplication. However, these methods require permuting the rows and columns of sparse matrices to generate more structured sparsity patterns. Martone et al.~\cite{martone2010blas, martone2014efficient} introduced the Recursive Sparse Block~(RSB) storage scheme for efficient execution of symmetric SpMV operations. On multi-core CPUs, this hierarchical representation demonstrates comparable performance to tiled (block-based) formats, such as Compressed Sparse Blocks (CSB)~\cite{bulucc2009parallel}. 
The Distributed Block Compressed Sparse Row~(DBCSR) library~\cite{borvstnik2014sparse} adopts a hybrid approach, in which a sparse matrix is recursively divided until it has a predefined number of blocks. Moreover, the DBCSR library provides a tensor interface~\cite{sivkov2019dbcsr} to execute tensor contractions using the optimized sparse matrix multiplication kernels.

\vspace*{-5pt}
\section{Conclusion} \label{sec:conclusion}
Motivated by the vulnerability of existing compressed formats to the irregular characteristics of higher-order sparse data, this work proposed the \ALTO format to efficiently encode sparse tensors of arbitrary dimensionality and spatial distributions using a single mode-agnostic representation. An implementation of the \ALTO-based tensor decomposition operations (all-modes MTTKRP) proved to outperform an oracle that selects the best state-of-the-art format, in terms of parallel performance and tensor storage. Specifically, \ALTO delivers $8\times$ geometric-mean speedup and $4.3\times$ geometric-mean compression ratio over the best general and mode-specific formats, respectively, due to its superior workload balance, adaptive synchronization, and compact encoding. 
Furthermore, the experiments showed the potential of our template-based implementation to bring extra performance improvement ($1.4\times$ geometric-mean speedup) by specializing the \ALTO-based tensor kernels for target application features. 

Our future work will investigate additional massively parallel and distributed-memory platforms that can benefit from the superior performance characteristics of \ALTO. We also plan to explore the use of the proposed format to accelerate other common sparse tensor algorithms, besides tensor decomposition.
\balance

\begin{acks}
The authors would like to thank Dr. Tamara G. Kolda for setting us on the direction of tensor linearzation and for providing valuable feedback.
\end{acks}

\bibliographystyle{ACM-Reference-Format}
\bibliography{alto}

\end{document}